%% file: main.tex
\documentclass{vldb}

\usepackage{wrapfig}
\usepackage{thmtools,thm-restate}
\usepackage{caption}
\usepackage{verbatim}
\usepackage{framed}
\usepackage[hidelinks]{hyperref}

\input{commands}

\lstdefinelanguage{scala}{
	morekeywords={abstract,case,catch,class,def,%
		do,else,extends,false,final,finally,%
		for,if,implicit,import,match,mixin,%
		new,null,object,override,package,%
		private,protected,requires,return,sealed,%
		super,this,throw,trait,true,try,%
		type,val,var,while,with,yield},
	otherkeywords={=>,<-,<\%,<:,>:,\#,@},
	sensitive=true,
	morecomment=[l]{//},
	morecomment=[n]{/*}{*/},
	morestring=[b]",
	morestring=[b]',
	morestring=[b]"""
}
\lstdefinelanguage{rheemlatin}{
	morekeywords={import,load,map,sample,repeat,reduce,reduceby,broadcast,key,store,with,using,platform,flatmap,filter,as,for,delimiter},
	otherkeywords={=>,<-,<\%,<:,>:,\#,@},
	keywordstyle=\color{javapurple}\bfseries,
	sensitive=true,
	morecomment=[l]{//},
	morecomment=[n]{/*}{*/},
	morestring=[b]",
	morestring=[b]',
	morestring=[b]"""
}

\begin{document}

\newcommand{\rheem}{\textsc{Rheem}\xspace}
\newcommand{\rheemx}{the \textsc{CPS} optimizer\xspace}
\newcommand{\Rheemx}{The \textsc{CPS} optimizer\xspace}
\newcommand{\open}{boundary\xspace}

\newcommand{\op}[1]{\protect\ensuremath{\mathsf{#1}}\xspace}
\newcommand{\task}[1]{\protect{\texttt{#1}}\xspace}
\newcommand{\pl}[1]{\protect\ensuremath{\mathsf{#1}}\xspace}
\newcommand{\ds}[1]{\protect{{\em #1}}\xspace}

\title{Building your Cross-Platform Application with RHEEM}

\numberofauthors{1}
\author{
	\vspace{0.1cm}
	Sanjay Chawla$^1$ \hspace{.5ex}
	Bertty Contreras-Rojas$^1$ \hspace{.5ex}
	Zoi Kaoudi$^1$ \hspace{.5ex}\\
	Sebastian Kruse$^{2}$\thanks{Work partially done while interning at QCRI.} \hspace{.5ex}
	Jorge-Arnulfo Quian\'e-Ruiz$^1$ \hspace{.5ex}
\vspace{0.3cm}\\
\affaddr{
	$^{1}${Qatar Computing Research Institute, Hamad Bin Khalifa University}
}\\
\vspace{1.5mm}
\affaddr{
	$^{2}${Hasso Plattner Institute, University of Potsdam}
}
\vspace{1.3mm}
\\
{
{\tt \url{http://da.qcri.org/rheem/}}
}
}

\maketitle

\begin{abstract}
Today, organizations typically perform tedious and costly tasks to juggle their code and data across different data processing platforms.
Addressing this pain and achieving automatic cross-platform data processing is quite challenging because it requires quite good expertise for all the available data processing platforms.
In this report, we present \rheem, a general-purpose cross-platform data processing system that alleviates users from the pain of finding the most efficient data processing platform for a given task.
It also splits a task into subtasks and assigns each subtask to a specific platform to minimize the overall cost (\eg~runtime or monetary cost).
To offer cross-platform functionality, it features
(i)~a robust interface to easily compose data analytic tasks;
(ii)~a novel cost-based optimizer able to find the most efficient platform in almost all cases; and
(iii)~an executor to efficiently orchestrate tasks over different platforms.
As a result, it allows users to focus on the business logic of their applications rather than on the mechanics of how to compose and execute them.
\rheem is released under an open source license.
\end{abstract}

\section{Introduction}
\label{section:intro}
\input{introduction}

\section{Cross-Platform Processing}
\label{section:crossplatform}
\input{crossplatform}

\section{Rheem Model}
\label{section:rheem}
\input{model}

\section{Rheem Internals}
\label{section:core}
\input{core}

\section{Rheem Interfaces}
\label{section:interfaces}
\input{interfaces}

\section{Examples of Rheem Plans}
\label{section:examples}
\input{examples}

\section{Rheem vs. Musketeer}
\label{section:musketeer}
\input{musketeer}

\section{Limitations}
\label{section:limitations}
\input{limitations}

\section{Related Work} 
\label{section:relatedwork}
\input{relatedwork}

\section{Conclusion}
\label{section:conclusion}
\input{conclusion}

\bibliographystyle{abbrv}
\bibliography{rheem}

\balance

%
%
%
%
%

\end{document}

%% file: commands.tex
\newcommand{\eat}[1]{}

\usepackage{balance}
\usepackage{latexsym}
\usepackage{amsfonts}
\usepackage{amsmath}
\usepackage{amssymb}
\usepackage{color}
\usepackage{colortbl}
\usepackage{epsfig}
\usepackage{xspace}
\usepackage{graphicx}
\usepackage{subfigure}
\usepackage{enumerate}
\usepackage[table]{xcolor}
\usepackage[all]{xy}
\usepackage{cite}
\usepackage{booktabs}
\usepackage{pdfpages}
\usepackage{pifont}

\usepackage{epsfig}
\usepackage{multirow}
\usepackage{url}

\newcommand{\at}[1]{\protect\ensuremath{\mathsf{#1}}\xspace}

\DeclareMathOperator*{\argmin}{arg\,min}

\newcommand{\bi}{\begin{itemize}}
\newcommand{\ei}{\end{itemize}}

\newcommand{\be}{\begin{enumerate}}
\newcommand{\ee}{\end{enumerate}}
\newcommand{\beqn}{\begin{eqnarray*}}
\newcommand{\eeqn}{\end{eqnarray*}}

\newcommand{\ie}{i.e.,\xspace}
\newcommand{\eg}{e.g.,\xspace}

\newcommand{\While}{\mbox{\bf while}\ }

\newcounter{ccc}

\newcommand{\eop}{\hspace*{\fill}\mbox{$\Box$}\vspace{1ex}}     

\newtheorem{example}{Example}

\newcounter{alg}[section]
\renewcommand{\thealg}{\nthesection.\arabic{alg}}

\newcounter{arule}
\renewcommand{\thearule}{\arabic{arule}}

\newcounter{claim}
\renewcommand{\theclaim}{\arabic{claim}}

\newcommand{\myparagraph}[1]{\vspace{0.1cm}\noindent\textbf{{#1}.~}}

\usepackage{soul}
\newif\ifnip

\newcounter{enum}
\newenvironment{packed_enum}{
\begin{list}{\textbf{(\arabic{enum})}}{
  \setlength{\itemsep}{0pt}
  \setlength{\parskip}{0pt}
  \setlength{\labelwidth}{-5 pt}
  \setlength{\leftmargin}{0 pt}
  \setlength{\itemindent}{0pt}
  \usecounter{enum}}
}{\end{list}}

\usepackage{listings}

\definecolor{javared}{rgb}{0.6,0,0} 
\definecolor{javagreen}{rgb}{0.25,0.5,0.35} 
\definecolor{javapurple}{rgb}{0.5,0,0.35} 
\definecolor{javadocblue}{rgb}{0.25,0.35,0.75} 

\usepackage[linesnumbered,ruled,vlined]{algorithm2e}
\let\oldnl\nl
\newcommand{\nonl}{\renewcommand{\nl}{\let\nl\oldnl}}

\makeatletter
    \newcommand\figcaption{\def\@captype{figure}\caption}
    \newcommand\tabcaption{\def\@captype{table}\caption}
\makeatother

%% file: introduction.tex

The pursuit of comprehensive, efficient, and scalable data analytics as well as the {\em one-size-does-not-fit-all} dictum have given rise to a plethora of data processing platforms ({\em platforms} for short).
These specialized platforms include DBMS, NoSQL, and MapReduce-like platforms.
In fact, just under the umbrella of NoSQL, there are reportedly over $200$ different platforms\footnote{\url{http://db-engines.com}}.
Each excels in specific aspects allowing applications to achieve high performance and scalability.
For example, while Spark supports \task{Select} queries, Postgres can execute them much faster by using indices.
However, Postgres is not as good as Spark for general purpose batch processing where parallel full scans are the key performance factor.
Several studies have shown this kind of performance differences~\cite{comparison,gog2015musketeer,thecase,systemml,ml4all}.

Moreover, today's data analytics is moving beyond the limits of a single platform.
For example:
(i)~IBM reported that North York hospital needs to process $50$ diverse datasets, which run on a dozen different platforms~\cite{ibm-health};
(ii)~Airlines need to analyze large datasets, which are produced by different departments, are of different data formats, and reside on multiple data sources, to produce global reports for decision makers~\cite{Fortunenews};
(iii)~Oil~\&~Gas companies need to process large amounts of diverse data spanning various platforms~\cite{baaziz2014,OilGas2013};
(iv)~Several data warehouse applications require data to be moved from a MapReduce-like system into a DBMS for further analysis~\cite{polybase,oraclehadoop}; and
(v)~Using multiple platforms for machine learning improves performance significantly~\cite{systemml,ml4all}.

To cope with these new requirements, developers (or data scientists) have to write ad-hoc programs and scripts to integrate different platforms.
This is not only a tedious, time-consuming, and costly task, but it also requires knowledge of the intricacies of the different platforms to achieve high efficiency and scalability.
Some systems have appeared with the goal of facilitating platform integration~\cite{apacheDrill,apacheFlume,luigi,prestoDB}.
Nonetheless, they all require a good deal of expertise from developers, who still need to decide which processing platforms to use for each task at hand.
Recent research has taken steps towards transparent cross-platform execution~\cite{SimitsisWCD12,polystores-blog,dbms+,gog2015musketeer,tensorflow2015-whitepaper,ires-bigdata}, but lacks several important aspects.
Usually these efforts do not automatically map tasks to platforms. Additionally, they do not consider complex data movement (\ie~with data transformations) among platforms~\cite{gog2015musketeer,ires-bigdata}. 
Finally, most of the research focuses on specific applications~\cite{SimitsisWCD12,dbms+,tensorflow2015-whitepaper}.

Therefore, there is a clear need for a systematic approach to enable efficient {\em cross-platform data processing}, \ie~use of multiple data processing platforms.
The Holy Grail would be to replicate the success of DBMSs for cross-platform data processing.
Users simply send their tasks expressing the logic of their applications, and the cross-platform system decides on which platform(s) to execute each task with the goal of minimizing its cost (\eg~runtime or monetary cost).
In other words, users focus on the high level details and the cross-platform system takes care of the low level details.

Building a cross-platform system is challenging on numerous fronts:
(i)~a cross-platform system not only has to effectively find all the suitable platforms for a given task, but also has to choose the most efficient one;
(ii)~cross-platform settings are characterized by high uncertainty as different platforms are autonomous and thus one has little control over them;
(iii)~the performance gains of using multiple platforms should compensate the added cost of moving data across platforms;
(iv)~it is crucial to achieve inter-platform parallelism to prevent slow platforms from dominating execution time; and
(v)~the system should be extensible to new platforms and application requirements.

\begin{figure}[t]
	\centering
	\includegraphics[scale=0.18]{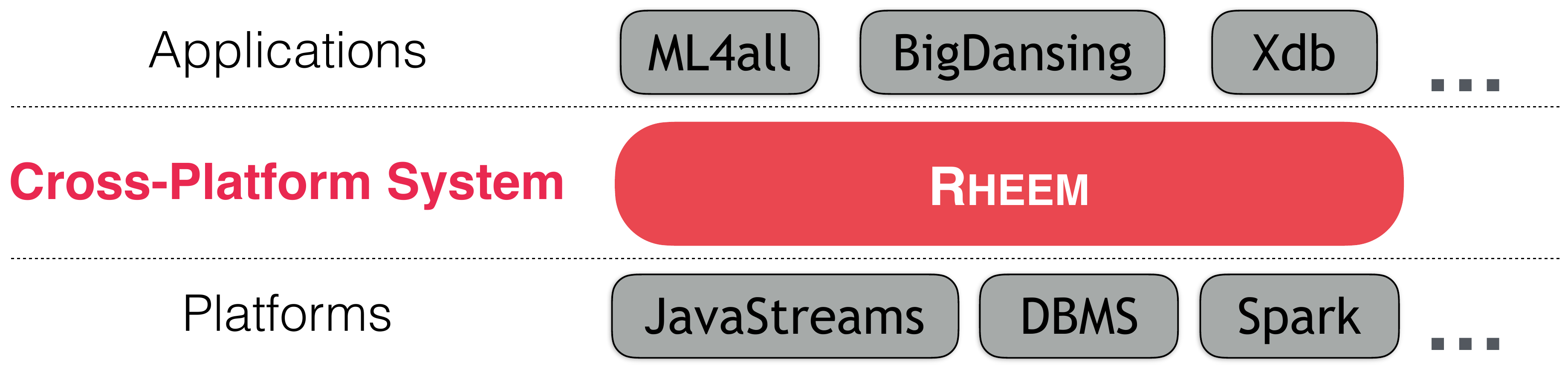}
	\vspace{-0.1cm}
	\caption{\rheem in the data analytics stack.}
	\label{figure:stack}
	\vspace{-0.6cm}
\end{figure}

In this report, we present \rheem\footnote{\rheem is open source under the Apache Software License~2.0 and can be found at \url{https://github.com/rheem-ecosystem/rheem}.}, the first general-purpose cross-platform system to tackle all of the above challenges.
The goal of \rheem is to {\em enable} applications and users to run data analytic tasks {\em efficiently} on one or more data processing platforms.
To do so, it decouples applications from platforms as shown in Figure~\ref{figure:stack}. 
Applications issue their tasks to \rheem, which in turn decides where to execute them.
As of today, \rheem supports a variety of platforms: \pl{Spark}, \pl{Flink}, \pl{JavaStreams}, \pl{Postgres}, \pl{GraphX}, \pl{GraphChi}, and \pl{Giraph}. 
We are currently testing \rheem in a large international airline company and in a biomedical research institute.
In the former case, we aim at seamlessly integrating all data analytic activity governing an aircraft;
In the latter case, we aim at reducing the effort scientists need for building data analytic pipelines while at the same time speeding up the running time.
Several papers show different aspects of \rheem: the vision behind it~\cite{rheem-vision}; its optimizer~\cite{rheem-tr}; its inequality join algorithm~\cite{iejoin}; and a couple of its applications~\cite{ml4all,bigdansing}.
A couple of demo papers showcase the benefits of \rheem~\cite{rheem-demo} and its interface~\cite{rheemstudio-demo}.
This report aims at presenting the complete design of \rheem and how all its pieces work together.

In summary, we identify four situations in which applications require support for cross-platform data processing in Section~\ref{section:crossplatform}.
For each case, we use a real application to show experimentally the benefits of cross-platform data processing using \rheem.
In Section~\ref{section:rheem}, we present the data and processing model of \rheem and show how it shields users from the intricacies of the underlying platforms.
\rheem provides flexible operator mappings that allow for better exploiting the underlying platforms. Also, its extensible design allows users to add new platforms and operators with very little effort.	
Then, in Section~\ref{section:core}, we discuss the key components of \rheem that make it novel:
among them a cost-based cross-platform optimizer that considers data movement costs;
a progressive optimization mechanism to deal with inconsistent cardinality estimates; and
a learning tool that alleviates users from the burden of tuning the cost model.
We present the \rheem interfaces whereby users can easily code and run a data analytic task in Section~\ref{section:interfaces}.
In particular, we present a data-flow language (RheemLatin) and a visual integrated development environment (\rheem Studio).
In Section~\ref{section:examples}, we show in detail three examples of real \rheem plans to better illustrate how developers can build their applications using these interfaces.
Section~\ref{section:limitations} outlines the limitations of \rheem.
Finally, we discuss related work in Section~\ref{section:relatedwork} and conclude with some open problems in Section~\ref{section:conclusion}.

%% file: crossplatform.tex
\begin{figure*}[t]
	\centering
	\includegraphics[scale=0.24]{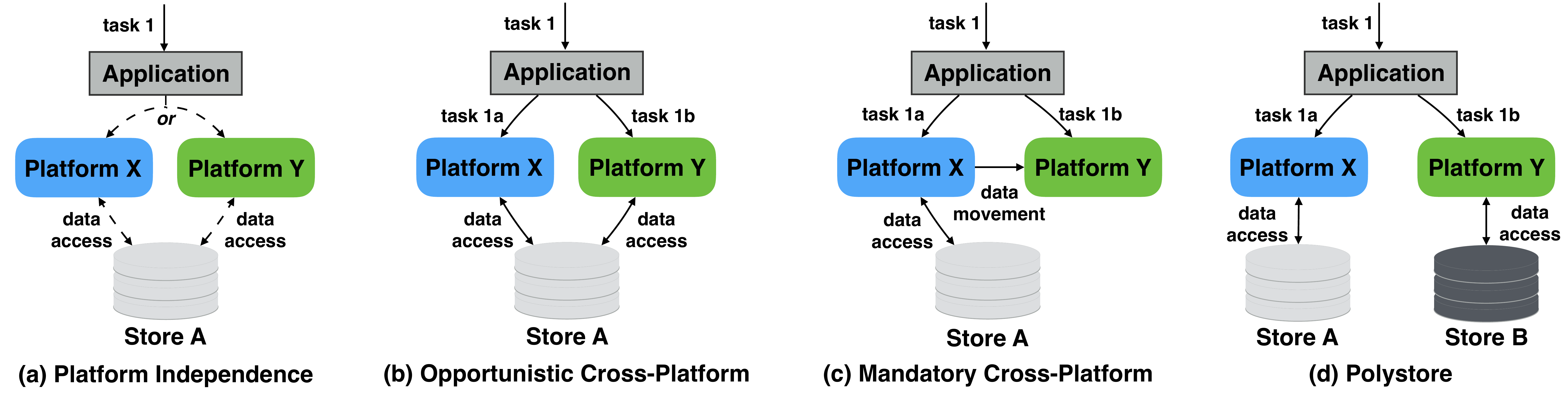}
	\vspace{-0.2cm}
	\caption{Cross-platform cases.}
	\label{figure:usecases}
\end{figure*}

We identified four situations in which an application requires support for cross-platform data processing~\cite{rheem-tutorial}. Figure~\ref{figure:usecases} illustrates these four cases.

 \begin{packed_enum}
\item {\em Platform-independence}. 
Applications run an entire task on a single platform but may require switching platforms for different input datasets or tasks usually with the goal of achieving better performance (Figure~\ref{figure:usecases}(a)).

\item {\em Opportunistic cross-platform}. 
Applications might also benefit performance-wise from using multiple platforms to run one single task (Figure~\ref{figure:usecases}(b)).

\item {\em Mandatory cross-platform}.
Applications may require multiple platforms because the platform where the input data resides, \eg~PostgreSQL, cannot perform the incoming task, \eg~a machine learning task. Thus, data should be moved from the platform it resides to another platform (Figure~\ref{figure:usecases}(c)).

\item {\em Polystore}.
Applications may require multiple platforms because the input data is stored on multiple data stores (Figure~\ref{figure:usecases}(d)).
\end{packed_enum}

In contrast to existing systems~\cite{gog2015musketeer,SimitsisWCD12,ires-bigdata,bigdawg-demo,myria}, \rheem helps users in {\em all} above cases.
The design of our system has been mainly driven by four applications:
a data cleaning application, {\em BigDansing}~\cite{bigdansing};
a machine learning application, {\em ML4all}~\cite{ml4all};
a database application, {\em xDB}; and
an end-to-end data discovery and preparation application, {\em Data Civilizer}~\cite{datacivilizerdemo}.
We use these applications to showcase the benefits of performing cross-platform data processing, instead of single-platform data processing, in terms of both performance and ease of use.

\begin{figure*}[t!]
	\centering
	\includegraphics[scale=0.35]{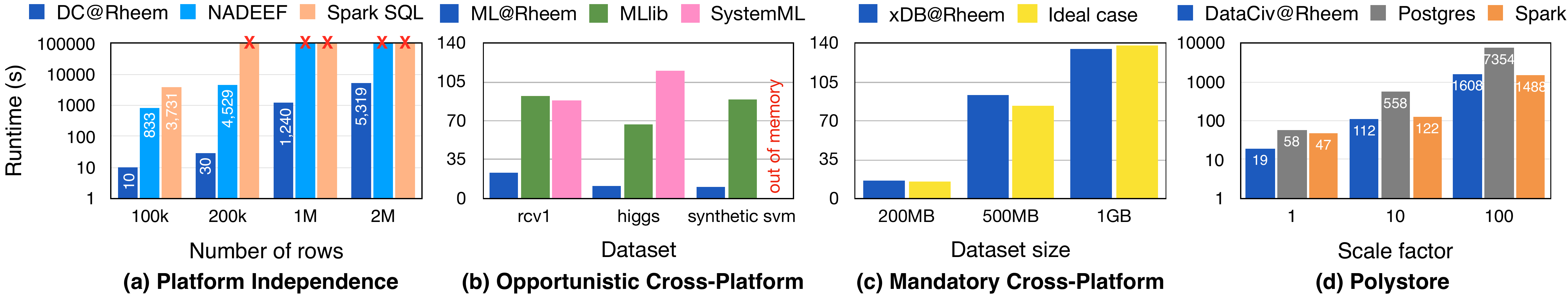}
	\vspace{-0.6cm}
	\caption{Benefits of the cross-platform data processing approach (using \rheem).}
	\label{figure:cross-platform}
	\vspace{-0.2cm}
\end{figure*}

\subsection{Platform Independence}
\label{section:usecases_independence}
Applications are usually tied to a specific platform.
This may not constitute the ideal case for two reasons.
First, as more efficient platforms become available, developers need to re-implement existing applications on top of these new platforms.
For example, Spark SQL~\cite{sparksql} and MLlib~\cite{mllib} are the Spark counterparts of Hive~\cite{hive} and Mahout~\cite{mahout}.
Migrating an application from one platform to another is a time-consuming and costly task and hence it is not always a viable choice.
Second, for different inputs of a specific task, a different platform may be the most efficient one, so the best platform cannot be determined statically.
For instance, running a specific task on a big data platform for very large datasets is often a good choice, while single-node platforms with only little overhead costs are often a better choice for small datasets~\cite{systemml}.
Thus, enabling applications to seamlessly switch from one platform to another according to the input dataset and task is important.
\rheem~{\em dynamically determines the best platform to run an incoming task.}

\myparagraph{Benefits}
We use BigDansing~\cite{bigdansing} to demonstrate the benefits of providing platform independence.
Users specify a data cleaning task with five logical operators:
\at{Scope} (identifies relevant data),
\at{Block} (defines the group of data among which an error may occur),
\at{Iterate} (enumerates candidate errors),
\at{Detect} (determines whether a candidate error is indeed an error),
and \at{GenFix} (generates a set of possible repairs).
\rheem maps these operators to \rheem operators to decide the best underlying platform.
We show the power of supporting cross-platform data processing by running an error detection task on a widely used Tax dataset~\cite{FanGJK08}.
 The task is based on the denial constraint
$\forall t_1,t_2, \neg(t_1.\at{Salary}>t2.\at{Salary}$  $\wedge t_1.\at{Tax}<t_2.\at{Tax})$, which states that there is an inconsistency between two tuples representing two different persons if one earns a higher salary but pays a lower tax.
We considered NADEEF~\cite{DBLP:conf/sigmod/DallachiesaEEEIOT13}, a data cleaning tool, and SparkSQL, a general-purpose framework, as baselines and forced \rheem to use either \pl{Spark} or \pl{JavaStreams} per run.

Figure~\ref{figure:cross-platform}(a) shows the results\footnote{The red cross means we stopped the execution after 40 hrs.}.
Overall, we observe that \rheem (DC@Rheem) allows data cleaning tasks to scale up to large datasets and be at least three orders of magnitude faster than baselines. 
One order of magnitude gain comes from the ability of \rheem to automatically switch platforms.
\rheem used JavaStreams for small datasets speeding up the data cleaning task by avoiding \pl{Spark}'s overhead, while it used \pl{Spark} for the largest datasets.
Furthermore, in contrast to \pl{Spark SQL} that cannot process inequality joins efficiently, \rheem's extensibility
allowed us to plug in a more efficient inequality-join algorithm~\cite{iejoin}, thereby further improving over these baselines.
In a nutshell, BigDansing benefited from \rheem because of its ability to effectively switch platforms and because of its extensibility to easily plug optimized algorithms.
We demonstrated how BigDansing benefits from \rheem in~\cite{rheem-demo}.

\subsection{Opportunistic Cross-Platform}
\label{section:usecases_opportunistic}

While some applications can be executed on a single platform, there are cases where their performance would be sped up by using multiple platforms.
For instance, users can run a gradient descent algorithm, such as SGD, on top of \pl{Spark} relatively fast.
Still, we recently showed that mixing it with \pl{JavaStreams} significantly improves performance~\cite{ml4all}.
In fact, opportunistic cross-platform processing can be seen as the execution counter-part of \emph{polyglot persistence}~\cite{polyglot-book}, where different types of databases are combined to leverage their individual strengths.
However, developing such cross-platform applications is difficult: developers must know all the cases where it is beneficial to use multiple platforms and how exactly to use them.
These opportunities are often very hard (if not impossible) to spot.
Even worse, like in the platform independence case, they usually cannot be determined a priori.
\rheem~{\em finds and exploits opportunities of using multiple processing platforms.}

\myparagraph{Benefits}
Let us now take our machine learning application, {\em ML4all}~\cite{ml4all}, to showcase the benefits of using multiple platforms to perform one single task.
ML4all  abstracts three fundamental phases (namely preparation, processing, and convergence) found in most machine learning tasks via seven logical operators which are mapped to \rheem operators.
In the preparation phase, the dataset is prepared appropriately along with the necessary initialization of the algorithm ({\tt Transform} and {\tt Stage} operators).
The processing phase computes the gradient and updates the current estimate of the solution ({\tt Sample}, {\tt Compute}, and {\tt Update} operators)
while the convergence phase repeats the processing phase based on the number of iterations or other criteria ({\tt Loop} and {\tt Converge} operators).
We demonstrate the benefits of using \rheem with a classification task over three benchmark datasets, using Stochastic Gradient Descent (SGD).

Figure~\ref{figure:cross-platform}(b) shows the results.
We observe that, even though all systems use the same SGD algorithm,
\rheem allows this algorithm to run significantly faster than competing Spark-based systems.
This is because of two main reasons.
First, this comes from opportunistically running the \op{Compute}, \op{Update}, \op{Converge}, and \op{Loop} operators on \pl{JavaStreams}, thereby avoiding some of the \pl{Spark}'s overhead.
\rheem runs the rest of the operators on \pl{Spark}.
\pl{MLlib} and \pl{SystemML} do not avoid such overhead by purely using Spark for the entire algorithm.
Second, ML4all leverages \rheem's extensibility to plug an efficient sampling operator, resulting in significant speedups.
We demonstrated how ML4all further benefits from \rheem in~\cite{rheem-demo}.


\subsection{Mandatory Cross-Platform}
\label{section:usecases_mandatory}
There are cases where an application needs to go beyond the functionalities offered by the platform on which the data is stored.
For instance, a dataset is stored on a relational database and a user needs to perform a clustering task on particular attributes.
Doing so inside the relational database might simply be disastrous in terms of performance.
Thus, the user needs to move the projected data out of the relational database and, for example, put it on HDFS in order to use Apache Flink~\cite{apacheFlink}, which is known to be efficient for iterative tasks.
A similar situation occurs in complex data analytics applications with disparate subtasks.
As an example, an application that extracts a graph from a text corpus to perform subsequent graph analytics may
require using both a text and a graph analytics system.
The required integration of platforms is tedious, repetitive, and particularly error-prone.
Nowadays, developers write ad-hoc programs to move the data around and integrate different platforms.
\rheem~{\em not only selects the right platforms for each task but also moves the data if necessary at execution time}.

\myparagraph{Benefits}
We use xDB\footnote{\url{https://github.com/rheem-ecosystem/xdb}}, a system on top of \rheem with database functionalities, to demonstrate the benefits of performing cross-platform data processing for the above situation.
It provides a declarative language to compose data analytic tasks, while its optimizer produces a plan to be executed in \rheem.
We evaluate the benefits of \rheem with the cross-community pagerank\footnote{This task basically intersects two community-DBpedia datasets and runs pagerank on the resulting dataset.} task, which is not only hard to express in SQL but also inefficient to run on a DBMS.
Thus, it is important to move the computation to another platform.
In this experiment, the input datasets are on Postgres and \rheem moves the data into Spark.

Figure~\ref{figure:cross-platform}(c) shows the results.
As a baseline, we consider the ideal case where the data is on HDFS and \rheem simply uses either JavaStreams or Spark to run the tasks.
We observe that \rheem allows xDB (xDB@Rheem) to achieve similar performance with the ideal case in all the situations, while fully automating the process.
This is a remarkable result as \rheem needs to move data out of Postgres to perform the tasks, in contrast to the ideal case.


\subsection{Polystore}
\label{section:usecases_polystore}
In many organizations, data is collected in different formats and on heterogeneous storage platforms ({\em data lakes}).
Typically, a data lake comprises various DBMSs, document stores, key-value stores, graph databases, and pure file systems.
As most of these stores are tightly coupled with an execution engine, \eg~a DBMS, it is crucial to be able to run analytics over multiple platforms.
For this, users perform not only tedious, time-intensive, and costly data migration, but also complex integration tasks for analyzing the data.
\rheem~{\em shields the users from all these tedious tasks and allows them to instead focus on the logic of their applications}.

\myparagraph{Benefits}
A clear example that shows the benefits of cross-platform data processing in a polystore case is the Data Civilizer system\cite{datacivilizerdemo}. 
Data Civilizer is a big data management system for data discovery, extraction, and cleaning from data lakes in large enterprises~\cite{datacivilizer}.
It constructs a graph that expresses relationships among data existing in heterogeneous data sources.
Data Civilizer uses \rheem to perform complex tasks over information that spans multiple data storages.
We measure the efficiency of \rheem for these polystore tasks with TPC-H query 5. 
In this experiment, we assume that the data is stored in HDFS (LINEITEM and ORDERS), Postgres (CUSTOMER, REGION, and SUPPLIER), and a local file system (NATION).
Thus, this task performs join, groupby, and orderby operations across three different platforms.
In this scenario, the common practice is to move the data into the database to enact the queries inside the database~\cite{polybase,oraclehadoop}
or move the data entirely to HDFS and use Spark.
We consider these two practices as the baseline. 
For a fairer comparison, we also set the ``parallel query" and ``effective IO concurrency" features of Postgres to $4$.

Figure~\ref{figure:cross-platform}(d) shows the results. 
\rheem (DataCiv@Rheem) is significantly faster, namely up to $5\times$, than the current practice.
We observed that loading data into Postgres is already approximately $3\times$ slower than it takes \rheem to complete the entire task.
Even when discarding data migration times, \rheem can still perform quite similarly to the parallel version of Postgres.
The pure execution time in Postgres for scale factor 100 amounts to $1,541$ sec compared to $1,608$ sec for \rheem, which exploits Spark for data parallelism.
We also observe that \rheem has negligible overhead over the case where the developer writes ad-hoc scripts to move the data to HDFS for running the task on \pl{Spark}. In particular, \rheem is twice faster than Spark for scale factor $1$ because it moves less data from Postgres to Spark.

%% file: model.tex
First of all, let us emphasize that \rheem is \emph{not} yet another data processing platform.
On the contrary, it is designed to work between applications and platforms (as shown in Figure~\ref{figure:stack}), helping applications to choose the right platform(s) for a given task.
\rheem is the first general-purpose cross-platform system that shields users from the intricacies of the underlying platforms and let them focus only on the logic of their applications.
We define the \rheem data and processing models in the following.

\begin{figure}[t!]
	\centering
	\includegraphics[scale=0.17]{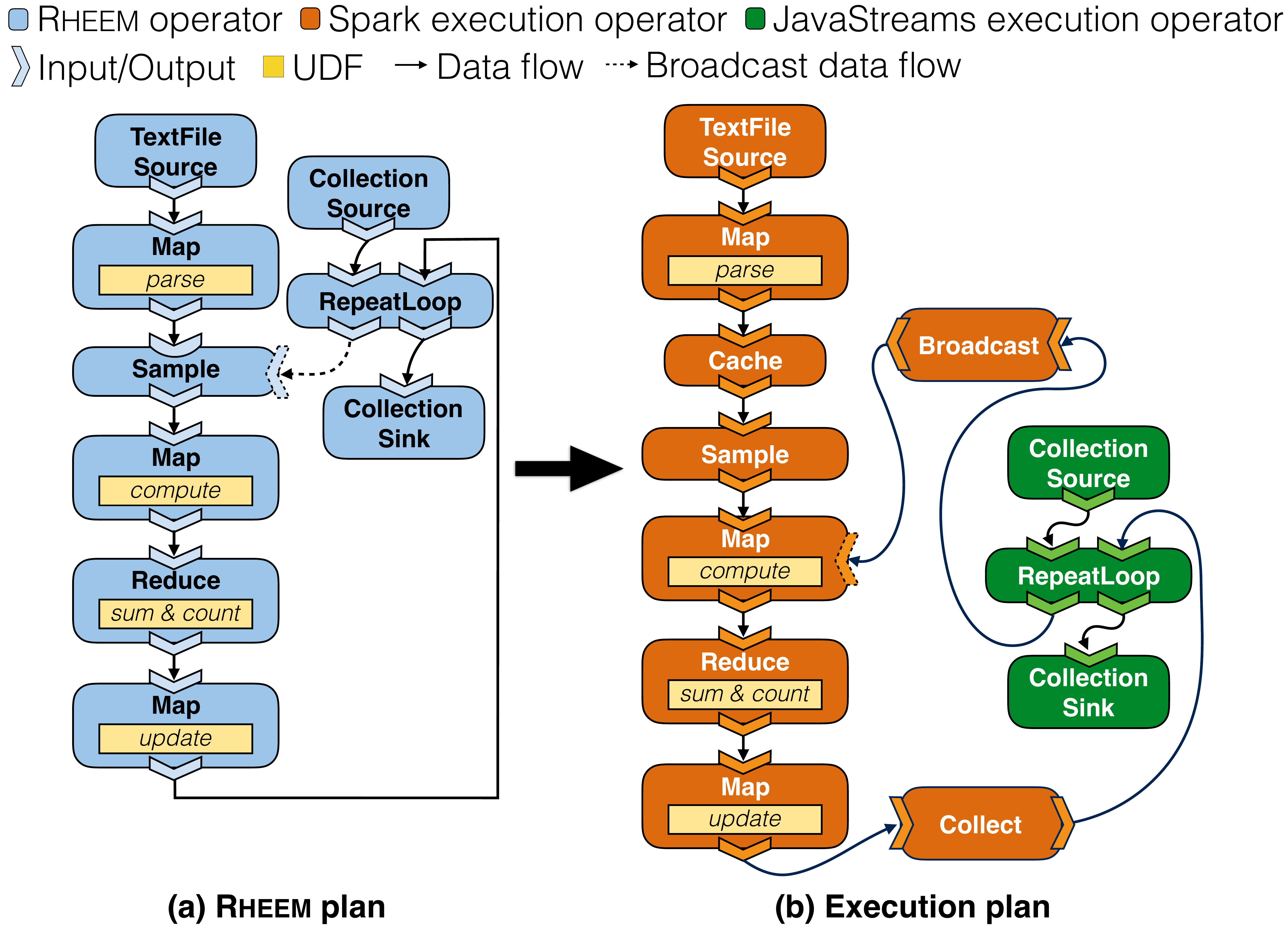}
	\vspace{-0.6cm}
	\caption{SGD example.}
	\label{figure:plan}
	\vspace{-0.5cm}
\end{figure}

\myparagraph{Data Quanta}
The \rheem data model relies on {\em data quanta}, the smallest processing units from the input datasets.
A data quantum 
can express a large spectrum of data formats, such as
database tuples, edges in a graph, or the full content of a document.
This flexibility allows applications and users to define a data quantum at any granularity level, \eg~at the attribute level rather than at the tuple level for a relational database.
This fine-grained data model allows \rheem to work in a highly parallel fashion, if necessary, to achieve better scalability and performance.

\myparagraph{\rheem Plan}
\rheem accepts as input a {\em \rheem plan}: a directed data flow graph whose vertices are {\em \rheem operators} and whose edges represent data flows among the operators.
A \rheem operator is a platform agnostic data transformation over its input data quanta, \eg~a \op{Map} operator transforms an individual data quantum while a \op{Reduce} operator aggregates input data quanta into a single output data quantum.
Only \op{Loop} operators accept feedback edges, which allows iterative data flows to be expressed.
Users or applications can refine the behavior of operators with a UDF.
Optionally, applications can also attach the selectivities of the operators through a UDF.
\rheem comes with default selectivity values in case they are not provided.
A \rheem plan must have at least one source operator, \ie~an operator reading or producing input data quanta, and one sink operator per branch, \ie~an operator retrieving or storing the result.
Intuitively, data quanta are flowing from source to sink operators, thereby being manipulated by all inner operators.
As our processing model is based on primitive operators, \rheem plans are highly expressive.
This is in contrast to other systems that accept either declarative queries~\cite{gog2015musketeer,myria} or coarse-granular operators~\cite{ires-bigdata}.

\vspace{-0.2cm}
\begin{example}
Figure~\ref{figure:plan}(a) shows a \rheem plan for the stochastic gradient descent algorithm (SGD).
Initially, the dataset containing the data points is read via a \op{TextFileSource} operator and parsed using a \op{Map} operator while the initial weights are read via a \op{Collection} source operator.
After the \op{RepeatLoop} operator, the weights are fed to the \op{Sample} operator, where a set of input data points is sampled.
Next, \op{Map(compute)} computes the gradient for each sampled data point.
Note that as \op{Map(compute)} requires all weights to compute the gradient, the weights are broadcasted at each iteration to the \op{Sample} operator (denoted by the dotted line).
Then, the \op{Reduce} operator computes the sum and count of all gradients.
The next \op{Map} operator uses these sum and count values to update the weights.
This process is repeated until the loop condition is satisfied.
The resulting weights are output in a collection sink.
\end{example}
\vspace{-0.2cm}

\myparagraph{Execution Plan}
Given a \rheem plan as input, \rheem uses a cost-based optimization approach to produce an {\em execution plan} by selecting one or more platforms to efficiently execute the input plan.
The cost can be any user-specified cost, \eg~runtime or monetary cost.
The resulting execution plan is again a data flow graph, where the vertices are now {\em execution operators}.
An execution operator implements one or more \rheem operators with platform-specific code.
For instance, the \op{Cache} Spark execution operator in \rheem implements the \op{Cache} \rheem operator by calling the \op{RDD.cache()} operation of Spark.
An execution plan may also comprise additional execution operators for data movement (\eg~data broadcasting) or data reuse (\eg~data caching).
Additionally, each execution operator has attached a UDF where its cost is specified.
\rheem learns such costs from execution logs using machine learning.
We discuss more details in Section~\ref{section:core_costlearner}.

\vspace{-0.2cm}
\begin{example}
Figure~\ref{figure:plan}(b) shows the SGD execution plan produced by \rheem when Spark and JavaStreams are the only available platforms.
This execution plan exploits high parallelism for the large dataset of input data points and avoids the extra overhead incurred by big data processing platforms for the smaller collection of weights.
Note that the execution plan also contains three execution operators for transferring (\op{Broadcast}, \op{Collect}) and making data quanta reusable across the platforms (\op{Cache}).
\end{example}
\vspace{-0.2cm}

\myparagraph{Operator Mappings}
To produce an execution plan, \rheem relies on flexible {\em m-to-n mappings} to map \rheem operators to execution operators.
Supporting $m$-to-$n$ mappings is particularly useful as it allows to map whole subplans of \rheem operators to subplans of execution operators.
Additionally, a subplan of \rheem (or execution) operators can map to another subplan of \rheem (respectively execution) operators.
As a result, we  can handle different abstraction levels among platforms, \eg~to emulate \rheem operators that are not natively supported by a specific platform.
This is not possible in other systems, such as~\cite{ires-bigdata}.

\begin{figure}[t!]
	\centering
	\includegraphics[scale=0.17]{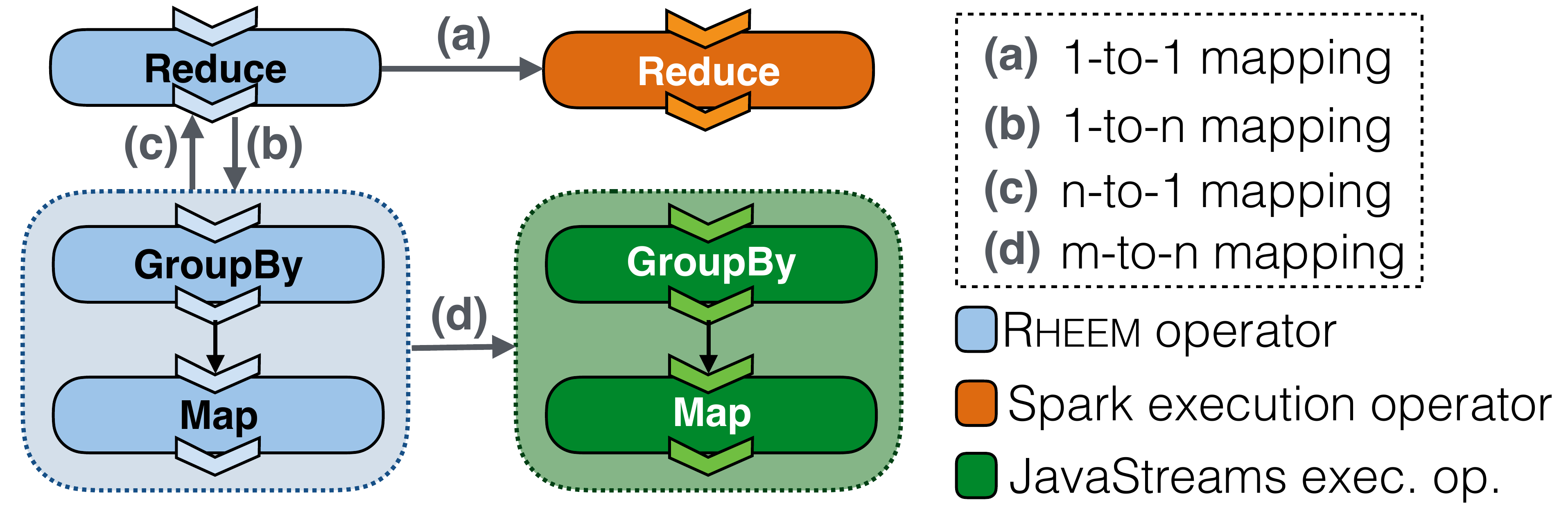}
	\vspace{-0.2cm}
	\caption{Operator mappings.}
	\label{figure:mappings}
	\vspace{-0.5cm}
\end{figure}

\vspace{-0.2cm}
\begin{example}
Figure~\ref{figure:mappings} illustrates the mapping for the \op{Reduce} \rheem operator.
This operator directly maps to the \op{Reduce} Spark execution operator via a 1-to-1 mapping (mapping~(a)).
However, it does not have a direct mapping to a JavaStreams execution operator.
Instead, it maps to a set of \rheem operators (\op{GroupBy} and \op{Map}) via a 1-to-n mapping (mapping~(b)) and vice-versa (n-to-1 mapping~(c)).
In turn, this set of \rheem operators maps to a set of JavaStreams execution operators (\op{GroupBy} and \op{Map}) via an m-to-n mapping (mapping~(d)).
\end{example}

\vspace{-0.2cm}
\myparagraph{Data movement}
Data flows among operators via {\em communication channels} (or simply {\em channels}).  A channel can be any internal data structure within a data processing platform (\eg~RDD for \pl{Spark} or Collection for \pl{JavaStreams}), or simply a file.
In the case of two execution operators of different platforms connected within a plan, it is necessary to convert the output channel of one to the input channel of the other (\eg~ from RDD to Collection). These conversions are handled by conversion operators, which in fact are regular execution operators.
For example, we can convert a Spark RDD channel to a JavaStreams Collection channel using the \pl{SparkCollect} operator (see Figure~\ref{figure:plan}(b)).
We represent the space of data movement paths across all platforms as a {\em channel conversion graph}, where the channels form its vertices and the {\em conversion operators} form its directed edges connecting one source channel to a target channel.
Unlike other approaches~\cite{ires-bigdata,gog2015musketeer}, developers do not need to provide conversion operators for all combinations of source and target channels.
It is thus much easier for developers to add new platforms to \rheem.

\myparagraph{Extensibility}
We designed \rheem to address extensibility as a first-class citizen rather than as ``nice-to-have'' feature.
Users add new \rheem and execution operators by merely extending or implementing few abstract classes/interfaces. 
\rheem provides template classes to facilitate the development for different operator types.
Users also add operator mappings by simply implementing an interface and specifying a graph pattern that matches the \rheem operator.
As a result, users can plug a new platform by providing:
(i)~its execution operators and their mappings; and 
(ii)~the communication channels that are specific to the new platform (\eg~\pl{RDDChannel} for Spark). 
Users neither have to modify the \rheem code nor integrate the newly added platform with all the already supported platforms.

%% file: core.tex

\begin{figure}[t!]
	\centering
	\includegraphics[scale=0.175]{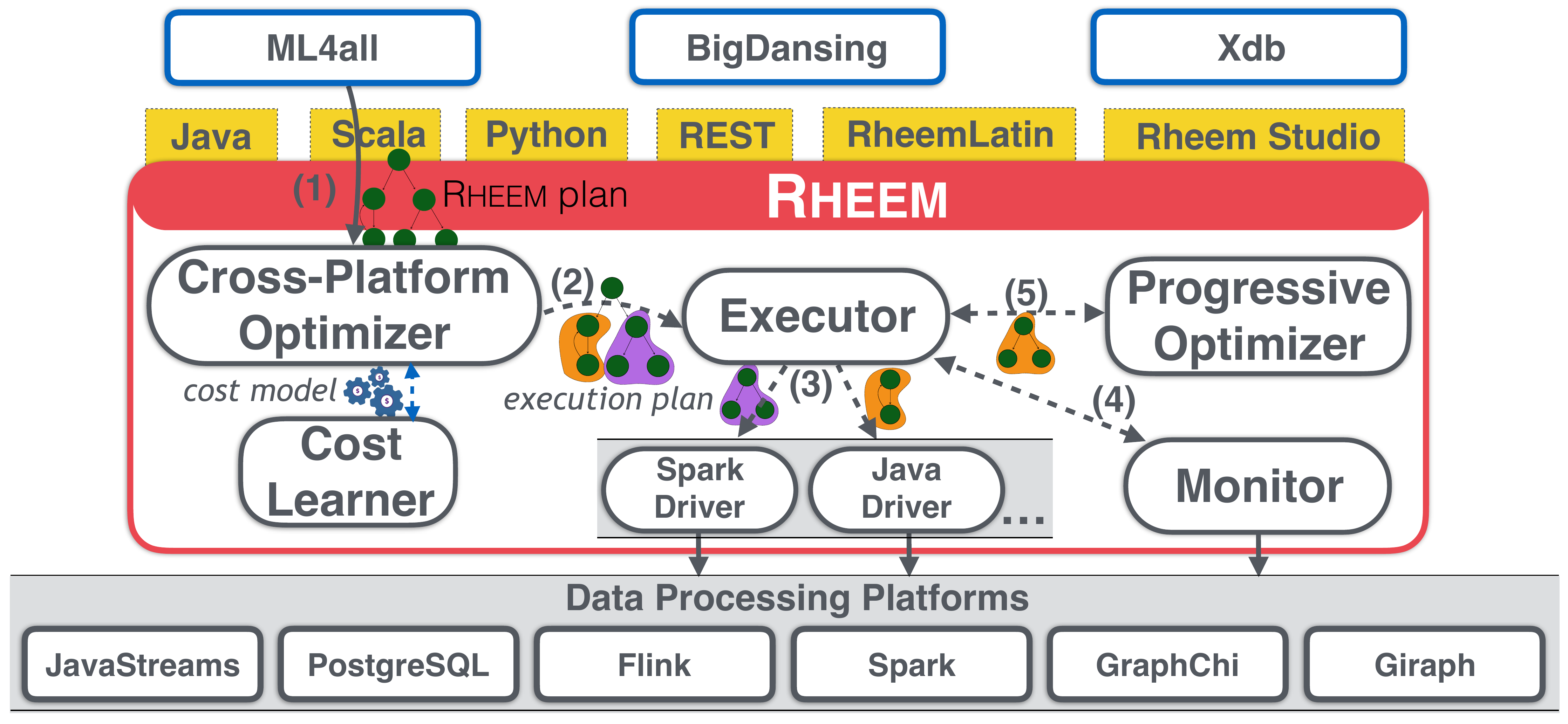}
	\vspace{-0.2cm}
	\caption{\rheem's ecosystem and architecture.}
	\label{figure:rheem}
	\vspace{-0.5cm}
\end{figure}

In this section, we give the details of the \rheem internals.
Figure~\ref{figure:rheem} depicts the \rheem ecosystem, \ie~the \rheem core architecture together with three main applications built on top of it.
Users provide a \rheem plan to the system (Step~(1) in Figure~\ref{figure:rheem}), using Java, Scala, Python, REST, RheemLatin, or \rheem Studio API (yellow boxes in Figure~\ref{figure:rheem}).
The {\em cross-platform optimizer} compiles the \rheem plan into an execution plan (Step~(2)), which specifies the processing platforms to use;
the {\em executor} schedules the resulting execution plan on the selected platforms (Step (3));
the {\em monitor} collects statistics and checks the health of the execution (Step (4));
the {\em progressive optimizer} re-optimizes the plan if the cardinality estimates turn out to be inaccurate (Step (5)); and
the {\em cost learner} helps users in building the cost model offline.
In the following, we explain each of these components using the pseudocode in
Algorithm~\ref{algo:optimization}, which shows the entire data processing pipeline.

\subsection{Optimizer}
\label{section:core_optimizer}
The cross-platform optimizer (Line~\ref{line:optimize} in Algorithm~\ref{algo:optimization}) is responsible for selecting the most efficient platform for executing each single operator in a \rheem plan.
One might think of a rule-based optimizer for selecting the right platforms to perform a given \rheem plan.
However, while a rule-based optimizer could determine how to split and execute a plan, \eg~based on its processing patterns~\cite{gog2015musketeer,myria}, it is neither practical nor effective.
First, by setting rules, one may make only very simplistic decisions based on the different cardinality and complexity of each operator.
Second, the cost of a task on any given platform depends on many input parameters, which hampers a rule-based optimizer's effectiveness as it oversimplifies the problem.
Third, as new platforms and applications emerge, maintaining a rule-based optimizer becomes cumbersome.

We thus pursue a more flexible cost-based approach:
{\em we split a given \rheem plan into subplans and determine the best platform for each subplan so that the total plan cost is minimized}.
Figure~\ref{figure:plan}(b) shows how the \rheem plan of Figure~\ref{figure:plan}(a) was split into two subplans to be executed in \pl{JavaStreams} and \pl{Spark}.
Below, we give the four main phases of the optimizer, namely {\em plan inflation}, {\em cardinality and cost estimation}, {\em data movement planning}, and {\em plan enumeration}.
Technical details about these can be found in~\cite{rheem-tr}.

\begin{algorithm} [t!]
	\footnotesize
	\caption{Cross-platform data processing\label{algo:optimization}}
	\KwIn{\rheem plan $\textit{rheemPlan}$}
	\BlankLine
	\DontPrintSemicolon
	$\textit{exPlan} \leftarrow$ $\texttt{Optimize}(\textit{rheemPlan})$\label{line:optimize}\;
	$\textit{monitor} \leftarrow \texttt{StartMonitor}(\textit{exPlan})$\label{line:monitor}\;
	$\textit{finished} \leftarrow \texttt{ExecuteUntilCheckpoint}(\textit{exPlan}, \textit{monitor})$\label{line:execution}\;
	\While{$\neg\text{finished}$}{
		$\textit{updated} \leftarrow \texttt{UpdateEstimates}(\textit{exPlan}, \textit{monitor})$\label{line:update}\;
		\lIf{$\textit{updated}$\label{line:proceed}}{
				$\textit{exPlan} \leftarrow \texttt{ReOptimize}(\textit{exPlan})$\label{line:reoptimize}
		}
		$\textit{finished} \leftarrow \texttt{ResumeExecution}(\textit{exPlan}, \textit{monitor})$\label{line:resume}\;
	}
\end{algorithm}

At first, the optimizer passes the \rheem plan through an inflation phase.
That is, it applies a set of operator mappings as described in Section~\ref{section:rheem}.
The optimizer then annotates the inflated plan with the cost of each execution operator.
\rheem represents cost estimates as intervals with a confidence value, which allows it to perform on-the-fly re-optimization as we will see in Section~\ref{section:core_progressive}.
The cost (\eg~wallclock time or monetary cost) of an execution operator depends on (i)~its resource usage (CPU, memory, disk, and network) and (ii)~the unit costs of each resource (\eg~how much one CPU cycle costs).
While the unit costs depend on hardware characteristics, the resource usage of each execution operator depends on its input cardinality.
Next, the optimizer looks for the best way to move data quanta among execution operators of different platforms.
As noted earlier, we model the problem of finding the most efficient communication path among execution operators as a graph problem, which we proved to be NP-hard.
Our solution to this problem relies kernelization and can discover all ways to connect execution operators of different platforms via a sequence of communication channels.
After the best data movement strategy is found, the optimizer attaches the data movement cost to the inflated plan.
At last, it determines the optimal way of executing a \rheem plan based on the cost estimates of its inflated plan.
For this, it must consider the previously computed data movement costs as well as the start-up costs of data processing platforms.
Thus, instead of taking a simple greedy approach that neglects data movement and platform start-up costs, we follow a principled approach: we use an enumeration algebra together with a lossless pruning technique.
Our pruning technique is guaranteed to not prune a subplan that is part of the optimal execution plan.
As a result, the optimizer can output the optimal execution plan without an exhaustive enumeration.

\subsection{Executor}
\label{section:core_executor}

The executor receives an execution plan from the optimizer to run it on the selected data processing platforms (Lines~\ref{line:execution} and \ref{line:resume} in Algorithm~\ref{algo:optimization}).
For example, the optimizer selected the Spark and JavaStreams platforms for our SGD example in Figure~\ref{figure:plan}(a).
Overall, the executor follows well-known approaches to parallelize a task over multiple compute nodes, with only few differences in the way it divides an execution plan.
In particular, it divides an execution plan into {\em stages}.
A stage is a subplan where
(i)~all its execution operators are from the same platform;
(ii)~at the end of its execution, the platforms need to give back the execution control to the executor; and
(iii)~its terminal operators materialize their output data quanta in a data structure, instead of being pipelined into the next operator.

\begin{figure}[t!]
	\centering
	\includegraphics[scale=0.175]{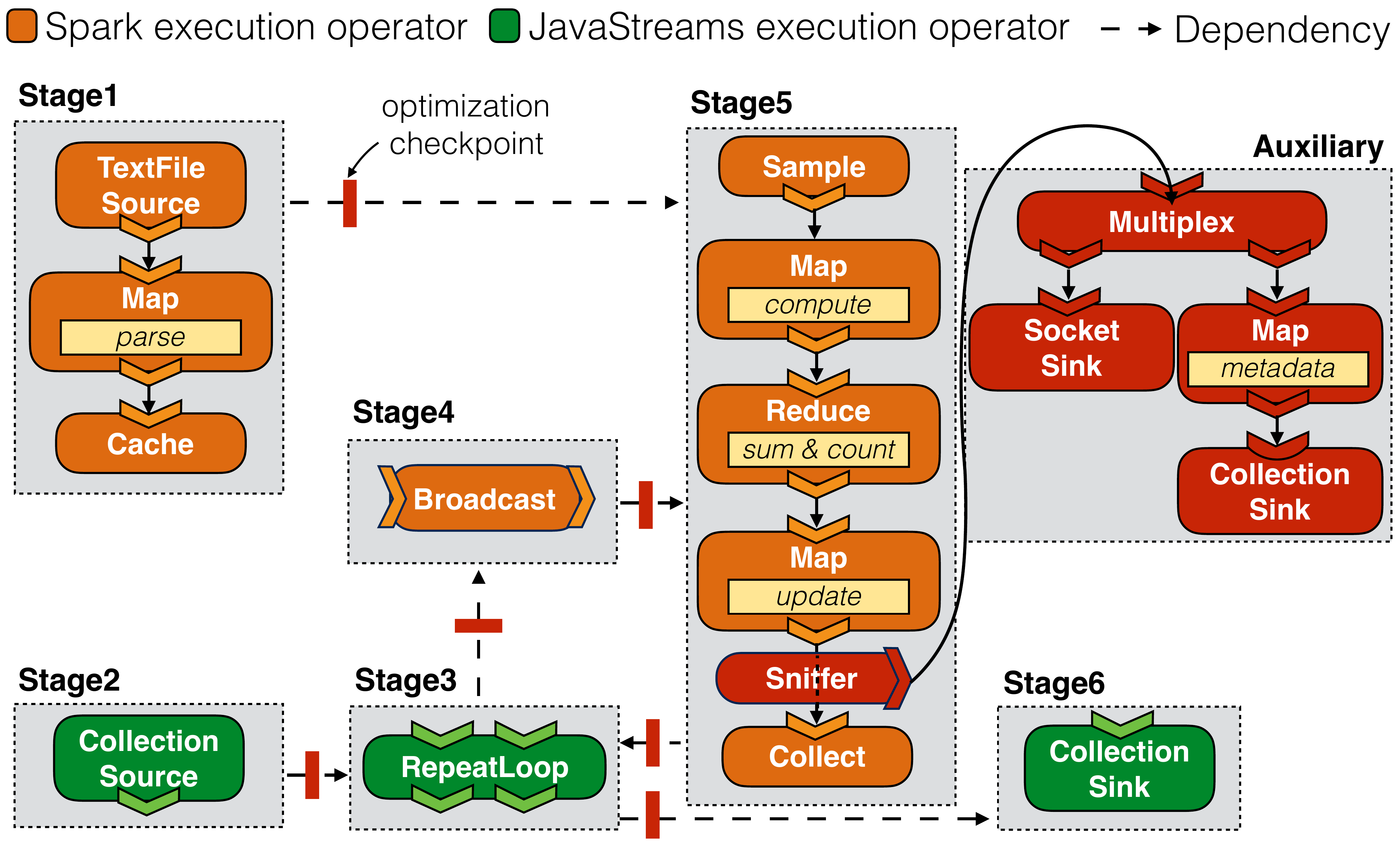}
	\caption{Stage dependencies for SGD.}
	\label{figure:dependencies}
	\vspace{-0.5cm}
\end{figure}

In our SGD example of Figure~\ref{figure:plan}(b), the executor divides the execution plan into six stages as illustrated in Figure~\ref{figure:dependencies}.
Note that Stage3 contains only the \op{RepeatLoop} operator as the executor must have the execution control to evaluate the loop condition.
This is why the executor also separates Stage1 from Stage5.
Then, it dispatches the stages to the relevant platform drivers, which in turn submit the stages as a job to the underlying platforms.
Stages are connected by data flow dependencies so that stages with no dependencies (\eg~Stage1 and Stage2) are dispatched first in parallel and any other stage is dispatched once its input dependencies are satisfied (\eg~Stage3 after Stage2).

\myparagraph{Data Exploration}
As data exploration is a key piece in the field of data science, the executor {\em optionally} allows applications to run in an {\em exploratory mode} where they can pause and resume the execution of a task at any point.
Achieving this in a cross-platform setting is very challenging, because most platforms, such as Spark, Flink, Giraph, Postgres, and Hadoop, do not support pausing task computations at all -- let alone resuming a task from an intermediate state.
Thus, the challenge resides in enabling the underlying platforms to support data exploration efficiently.
\rheem achieves this by injecting {\em sniffers} into execution plans and attaching {\em auxiliary} execution plans.
A sniffer is an execution operator that duplicates intermediate results and sends them to an auxiliary execution plan.
For example, the user would like to keep track of the weights at each iteration of SGD
and thus a sniffer is necessary right after updating the weights (Stage5 in Figure~\ref{figure:dependencies}).
The sniffer sends the weights to an auxiliary plan that is responsible for reporting them back to the user (the socket sink operator in Figure~\ref{figure:dependencies}).
This auxiliary plan is also responsible for computing and storing additional metadata
for efficient task resumption (the map and collection sink operators of the auxiliary plan in Figure~\ref{figure:dependencies}).
When resuming a task, the executor performs the task by re-using as much as possible from the previously computed metadata.
For instance, if the user pauses the SGD task at iteration $i$ and resumes it later on, the executor fetches the previously computed weights of iteration $i$ and resumes the task. 


\subsection{Monitor}
\label{section:core_monitor}
Recall that the cross-platform optimizer operates in a setting that is characterized by high uncertainty.
For instance, the semantics of UDFs and data distributions are usually unknown because of the little control over the underlying platforms.
This uncertainty can cause poor cardinality and cost estimates and hence can negatively impact the effectiveness of the optimizer~\cite{howgoodoptimizersare}.
To compensate this uncertainty, \rheem registers the execution of a plan with the monitor (Line~\ref{line:monitor} in Algorithm~\ref{algo:optimization}).
The monitor collects light-weight execution statistics for the given plan, such as data cardinalities and operator execution times.
It is also aware of lazy execution strategies used by the underlying platforms and assigns measured execution time correctly to operators.
\rheem uses these statistics to improve its cost model and re-optimize ongoing execution plans in case of poor cardinality estimates.
Additionally, the monitor is responsible for checking the health of the execution.
For instance, if it finds a large mismatch between the real output cardinalities and the estimated ones, it pauses the execution plan and sends it to the progressive optimizer.

\subsection{Progressive Optimizer}
\label{section:core_progressive}

To mitigate the effects of bad cardinality estimates, \rheem employs a \emph{progressive query optimization} approach.
The key principle is to re-optimize the plan whenever the cardinalities observed by the monitor greatly mismatch the estimated ones~\cite{markl04}.
Applying progressive query optimization in our setting comes with two main challenges.
First, we have only limited control over the underlying platforms, which makes plan instrumentation and halting executions difficult.
Second, re-optimizing an ongoing execution plan must efficiently consider the results already produced.

We tackle these challenges by using \emph{optimization checkpoints}.
An optimization checkpoint tells the executor to pause the plan execution in order to consider a re-optimization of the plan beyond the checkpoint.
The progressive optimizer inserts optimization checkpoints into execution plans wherever
(i)~cardinality estimates are uncertain (having a wide interval or low confidence) or
(ii)~the data is at rest (\eg~a Java collection or a file).
For instance, the optimizer inserts an optimization checkpoint right after Stage1 as 
the data is at rest because of the \op{Cache} operator (see Figure~\ref{figure:dependencies}).
When the executor cannot dispatch a new stage anymore without crossing an optimization checkpoint, it pauses the execution and gives the control to the progressive optimizer. 
The latter gets the actual cardinalities observed so far by the monitor and re-computes all cardinalities from the current optimization checkpoint (Line~\ref{line:update} in Algorithm~\ref{algo:optimization}).
In case of a mismatch, it re-optimizes the remaining of the plan (from the current optimization checkpoint) using the new cardinalities (Line~\ref{line:reoptimize}).
It then gives the new execution plan to the executor, which resumes the execution from the current optimization checkpoint (Line~\ref{line:resume}).
\rheem can switch between execution and progressive optimization any number of times at a negligible cost.

\subsection{Cost Model Learner}
\label{section:core_costlearner}
Profiling operators in isolation might be unrealistic whenever platforms optimize execution across multiple operators, \eg~by pipelining. 
Indeed, we found cost functions derived from isolated benchmarking to be insufficiently accurate.
We thus take a different approach.

\myparagraph{Learning the Cost Model}
Recall that each execution operator $o$ is associated with a number of resource usage functions ($r_o^m$, where $m$ is CPU, memory, disk, or network). 
For instance, the cost function to estimate the CPU cycles required by the \at{JavaFilter} operator is $r_{JavaFilter}^{CPU} := c_{in} \times (\alpha + \beta) + \delta$, where parameters $\alpha$ and $\beta$ denote the number of required CPU cycles for each input data quantum in the operator itself and in its UDF, and parameter $\delta$ describes some fixed overhead for the operator start-up and scheduling.
We then multiply each of these resource usage functions $r_o^{m}$ with the time required per unit (\eg~msec/CPU cycle) to get the time estimate $t_o^m$.
The total cost estimate for operator $o$ is defined as: $f_o = t_o^{CPU} + t_o^{mem} + t_o^{disk} + t_o^{net}$.
However, obtaining the parameters for each resource, such as the $\alpha, \beta, \delta$ values for CPU, is not trivial.
We, thus, use execution logs to {\em learn} these parameters in an offline fashion and model the cost of individual execution operators as a {\em regression problem}.
Note that the execution logs contain the runtimes of execution stages (\ie~pipelines of operators as defined in Section~\ref{section:core_executor}) and not of individual operators.
Let $(\{(o_1, C_1), (o_2, C_2), \dots (o_n, C_n)\}, t)$ be an execution stage, with $o_i$, $0< i \leq n$, where $o_i$ are execution operators, $C_i$ are input and output true cardinalities, and $t$ is the measured execution time for the entire stage.
Furthermore, let $f_{i}({\bf x}, C_i)$ be the total cost function for execution operator $o_i$ with ${\bf x}$ being a vector with the parameters of all resource usage functions (\eg~CPU cycles, disk I/O per data quantum).
We are interested in finding ${\bf x}_\text{min}=\argmin_{\bf x}\ loss\left(t,~\sum_{i=1}^n{f_i({\bf x}, C_i)}\right)$.
Specifically, we use a \emph{relative} loss function defined as $loss(t, t')= \left(\frac{|t - t'| + s}{t + s}\right)^2$,
\noindent where $t'$ is the geometric mean of the lower and upper bounds of the cost estimate produced by $\sum f_i({\bf x}, C_i)$ and $s$ is a regularizer inspired by additive smoothing that tempers the loss for small $t$.
Note that we can easily generalize this optimization problem to multiple execution stages:
we minimize the weighted arithmetic mean of the losses of multiple execution stages. In particular, we use as stage weights the sum of the relative frequencies of the stages' operators among all stages, so as to deal with skewed workloads that contain certain operators more often than others.
Finally, we apply a genetic algorithm~\cite{mitchell1998introduction} to find ${\bf x}_\text{min}$.
In contrast to other optimization algorithms, genetic algorithms impose only few restrictions on the loss function to be minimized.
Hence, our cost learner can deal with arbitrary cost functions.
Applying this technique allows us to calibrate the cost functions with only little additional effort.

\myparagraph{Logs Generation}
 Clearly, the more execution logs are available, the better \rheem can tune the cost model.
Thus, \rheem comes with a log generator. 
It first creates a set of \rheem plans by composing all possible combinations of \rheem operators forming a particular topology.
We found that most data analytic tasks in practice follow three different topologies: {\em pipeline} (\eg~batch tasks), {\em iterative} (\eg~ML tasks), and {\em merge} (\eg~SPJA tasks).
It then generates all possible executions plans for the previously created set of \rheem plans.
Next, it creates different configurations for each execution plan, \ie~it varies the UDF complexity, output cardinalities, input dataset sizes, and data types.
Once it has generated all possible plans with different configurations, it executes them and logs their runtime. 


%% file: interfaces.tex

\rheem provides a set of native APIs for developers to build their applications. 
These include Java, Scala, Python, and REST. Examples of using these APIs can be found in the \rheem repository\footnote{\small{\url{https://github.com/rheem-ecosystem/rheem-benchmark}}}.
The code developers have to write is fully agnostic of the underlying platforms.
Still, in case the user wants to force \rheem to execute a given operator on a specific platform, she can invoke the \textsf{\small withTargetPlatform} method.
Similarly, she can force the system to use a specific execution operator via the \textsf{\small customOperator} method, which further enables users to employ custom operators without having to extend the API.

Although the native APIs are quite popular among developers, many users are not proficient using these APIs.
Thus, \rheem also provides two APIs that target non-expert users: a data-flow language ({\em RheemLatin}) and a visual IDE ({\em {\sc Rheem} Studio}).
We explain these interfaces using our SGD example from Figure~\ref{figure:plan}.
However, for the sake of explanation, before going into the details of these two interfaces, we first show how one can implement SGD on \rheem using one of its native APIs.
The salient feature of all these APIs is that they are all platform-agnostic.
It is \rheem that figures out on which platform to execute each of the operators.

\subsection{Platform-Agnostic Native API}
\label{section:api_native}
Let us explain how users can code their applications using one of the native APIs of \rheem.
We use the Scala API and our SGD running example (see Listing~\ref{list:plan})\footnote{The complete source code of this task is available online: \url{https://github.com/rheem-ecosystem/rheem-benchmark}.}.
\begin{lstlisting}[basicstyle=\scriptsize\normalfont\sffamily,aboveskip=0pt,belowskip=0pt,float=h!,label=list:plan,numbers=none,numbers=left,caption=SGD task using the Scala API.,language=scala]
val context = new RheemContext(new Configuration)
	.withPlugin(Spark.basicPlugin)
	.withPlugin(JavaStreams.basicPlugin)
val plan = new PlanBuilder(context)
val points = plan.readTextFile("hdfs://myData.csv")
	.map(parsePoints)
val finalWeights = plan.loadCollection(createRandomWeights())
	.repeat(50, { weights =>
		points.sample(sampleSize).withBroadcast(weights)
		.map(computeGradient())
		.reduce(_ + _)
		.map(updateWeights())
	}).collect()
\end{lstlisting}

First, a user creates the \rheem context, where she specifies the available platforms (Lines~1-3): Spark and JavaStreams in this example.
She then initializes her \rheem plan with this context (Line~4).
Eventually, she creates the graph of \rheem operators that defines the SGD task (Lines~5-13).
Note that \rheem plans must have at least one source operator (Line~5), \ie~an operator reading or producing input data quanta, and one sink operator per branch (Line~13), \ie~an operator retrieving or storing the result.
Recall that a \rheem plan must have at least one source operator (Line~5) and one sink operator per branch (Line~13).
Also, observe that this code is fully agnostic of the underlying platforms.
Still, in case the user wants to force \rheem to execute a given operator on a specific platform, she can invoke the \textsf{\small withTargetPlatform} method.
Similarly, she can force the system to use a specific execution operator via the \textsf{\small customOperator} method, which further enables users to employ custom operators without having to extend the API.
For clarity reasons, we did not include the UDF implementations in Listing~\ref{list:plan}.

\subsection{RheemLatin}
\label{section:api_rheemlatin}
\rheem provides a data-flow language (RheemLatin) for users to specify their tasks~\cite{rheemstudio-demo}.
Our goal is to provide ease-of-use to users without compromising expressiveness.
RheemLatin follows a procedural programming style to naturally fit the pipeline paradigm of \rheem.
This is similar to the R language, which is quite popular among data scientists.
It draws its inspiration from PigLatin~\cite{piglatin} and hence it has PigLatin's grammar and supports most PigLatin's keywords.
In fact, one could see it as an extension of PigLatin for cross-platform settings.
For example, users can specify the platform for any part of their queries.
More importantly, it provides a set of configuration files whereby users can add new keywords to the language together with their mappings to \rheem operators.
As a result, users can easily adapt RheemLatin for their applications.
Listing~\ref{list:latin} illustrates how one can express our SGD example with RheemLatin.

\begin{lstlisting}[basicstyle=\scriptsize\normalfont\sffamily,aboveskip=0pt,belowskip=0pt,float=h!,label=list:latin,numbers=none,numbers=left,caption=SGD task in RheemLatin.,language=rheemlatin]
import '/sgd/udfs.class' as taggedPointCounter;
lines = load 'hdfs://myData.csv';
points = map lines -> {taggedPointCounter.parsePoints(lines)};
weights = load taggedPointCounter.createWeights();
final_weights = repeat 50 {
   sample_points = sample points -> {taggedPointCounter.getSample()} with broadcast weights;
   gradient = map sample_points -> {taggedPointCounter.computeGradient()};
   gradient_sum_count = reduce gradient -> {gradient.sumcount()};
   weights = map gradient_sum -> {gradient_sum_count.average()} with platform 'JavaStreams';}
store final_weights 'hdfs://output/sgd';
\end{lstlisting}

The user starts by importing all her required UDFs (Line~1).
She then parses all the data points from the input dataset (Lines~2 and~3) and initializes the weights (Line~4).
Next, she proceeds to perform the core of SGD: she takes a sample of data points (Line~6), computes the gradient for each sampled data point (Line~7), updates the weights (Lines~8 and~9), and repeats the process $50$ times (Line~5).
She can also repeat such a core process until convergence by using \textsf{WhileLoop} instead of \textsf{Repeat}.
Optionally, she can specify the platform for any part of her query.
For instance, she might know that updating the weights on each iteration is a lightweight computation and hence might specify to use JavaStreams (Line~9).
She finishes by storing the final weights on HDFS (Line~10).

\subsection{Rheem Studio}
\label{section:api_studio}
\begin{figure}[h!]
	\centering
	\includegraphics[scale=0.072]{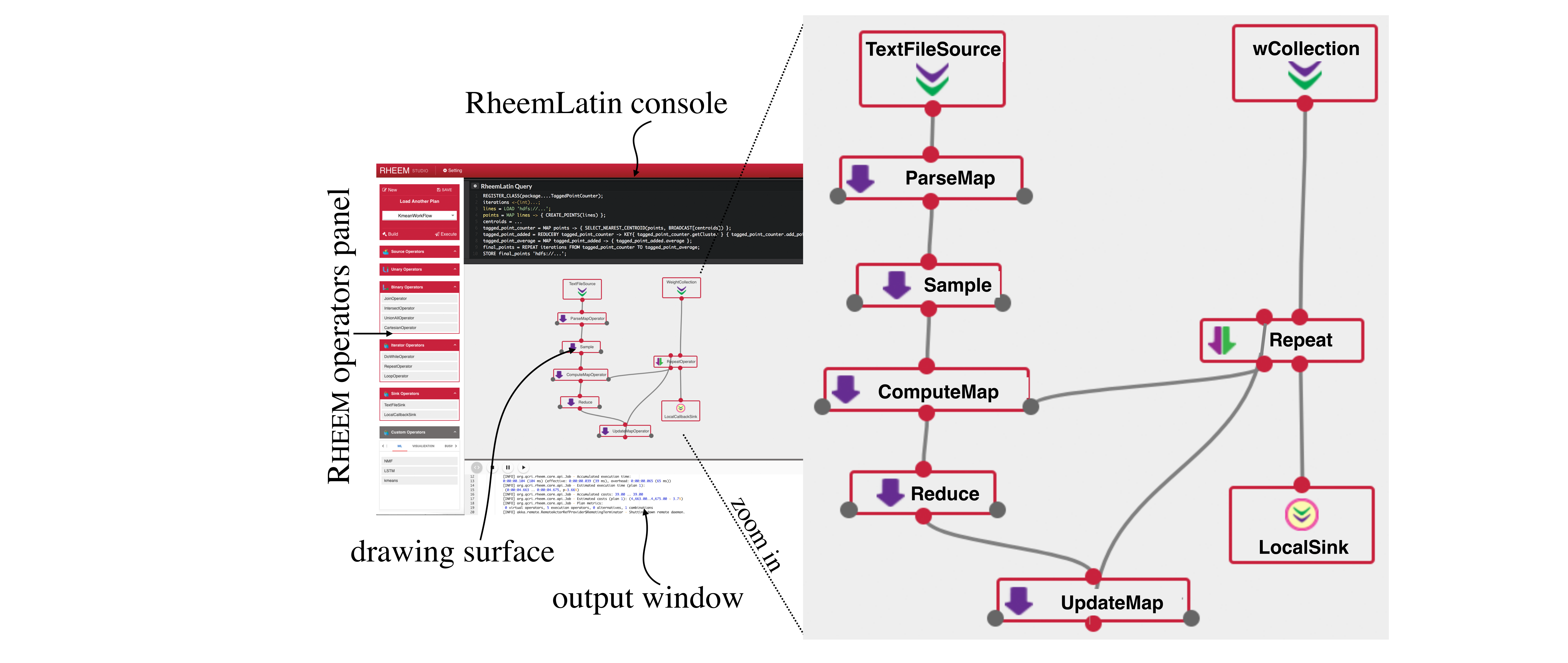}
	\vspace{-0.5cm}
	\caption{SGD task in the \rheem Studio.}
	\label{figure:rstudio}
	\vspace{-0.5cm}
\end{figure}

Although the native APIs and RheemLatin cover a large number of users, some might still be unfamiliar with programming and data-flow languages.
Also, some other users may simply desire to speed up the process of composing their data analytic tasks.
To this end, \rheem provides a visual IDE (\rheem Studio) where users can compose their data analytic tasks in a {\em drag and drop} fashion~\cite{rheemstudio-demo}.
Figure~\ref{figure:rstudio} shows the \rheem Studio's GUI.
The GUI is composed of four parts: a panel containing all \rheem operators, the drawing surface, a console for writing RheemLatin queries, and the output terminal.
The right-side of Figure~\ref{figure:rstudio} shows how operators are connected for an SGD plan.
The studio provides default implementations for any of the \rheem operators, which enables users to run common data analytic tasks without writing code.
Yet, expert users can provide a UDF by double-clicking on any operator.

Users can draw such a plan by simply {\em dragging} as many \rheem operators as required from the left-side panel and {\em dropping} them on the drawing surface.
They consequently connect the operators as required by their data analytic task.
The right-side of Figure~\ref{figure:rstudio} shows how operators are connected for SGD.
While connecting operators, the studio validates such connections and gives feedback to users in case that a connection cannot be established, \eg~the output and input of two connected operators are of different data types.
Last but not least, the studio provides default implementations for any of the \rheem operators, which enables users to run common data analytic tasks without writing a single line of code.
Yet, expert users can provide a UDF by double-clicking on any operator.

%% file: examples.tex

We now provide in detail three examples of how users can implement their tasks using the Scala native API and the RheemLatin interface.
For this, we consider three popular data analytic tasks: {\em WordCount} (a well-known aggregate task), {\em K-means} (a very representative iterative task), and {\em PolyJoin} (a common task over difference data sources).

Users start their \rheem plans in Scala with a preamble that defines the context and the platforms to be used, as shown in Listing~\ref{list:preamble}.
For the sake of presentation, we do not include this preamble in our Scala code examples below.

\begin{lstlisting}[basicstyle=\scriptsize\normalfont\sffamily,aboveskip=0pt,belowskip=0pt,float=h!,label=list:preamble,numbers=none,numbers=left,caption=Preamble in the Scala API.,language=scala]
val context = new RheemContext(new Configuration)
	.withPlugin(Spark.basicPlugin)
	.withPlugin(JavaStreams.basicPlugin)
val plan = new PlanBuilder(context)
\end{lstlisting}


{\em WordCount} is an aggregate task that computes the frequency with which each word appears in a dataset.
Listing~\ref{list:wclatin} shows the RheemLatin query for this task:
Line~1 imports all the required UDFs, Line~2 loads the input data;
Lines~3 and~4 parse the words and convert them into records;
Line~5 computes the frequency of each word; and
Line~6 stores the final word count on disk.
Note that users naturally define the flow of their analytical tasks with RheemLatin.
Alternatively, users can implement this task using one of the native APIs of \rheem.
Listing~\ref{list:wordcount} shows the Scala code for this task.
Similar to the RheemLatin query, the Scala code keeps the plan composition simple.

\begin{lstlisting}[basicstyle=\scriptsize\normalfont\sffamily,aboveskip=0pt,belowskip=0pt,float=h!,label=list:wclatin,numbers=none,numbers=left,caption=Word Count task in RheemLatin.,language=rheemlatin]
import '/wordcount/udfs.class' as wordcount;
lines = load 'hdfs://myWords.txt';
words = flatmap lines -> { wordcount.splitWords() };
tuples = map words -> { wordcount.convert2Tuple() };
adds   = reduce tuples -> { wordcount.getWord() }, tuples -> { wordcount.reduce() };
store adds '/output/wordcount'; 
\end{lstlisting}

\begin{lstlisting}[basicstyle=\scriptsize\normalfont\sffamily,aboveskip=0pt,belowskip=0pt,float=h!,label=list:wordcount,numbers=none,numbers=left,caption=Word Count task using the Scala API.,language=scala]
val words = plan.readTextFile("hdfs://myWords.csv")
	.flatMap(_.split("\\W+"))
	.map(word => (word.toLowerCase, 1))
	.reduceByKey(_._1, (c1, c2) => (c1._1, c1._2 + c2._2))
	.collect()
\end{lstlisting}


{\em K-means} is a widely used ML task for clustering data points together according to their similarity.
We show the RheemLatin query in Listing~\ref{list:kmeanslatin}.
In contrast to the WordCount task, this task is iterative (Lines~4--7).
We observe that defining loops in RheemLatin is quite similar to coding in a high-level language (\eg~Scala), which makes it intuitive for most users.
Listing~\ref{list:kmeanplan} shows its counterpart in Scala.

\begin{lstlisting}[basicstyle=\scriptsize\normalfont\sffamily,aboveskip=0pt,belowskip=0pt,float=h!,label=list:kmeanslatin,numbers=none,numbers=left,caption=K-means task in RheemLatin.,language=rheemlatin]
lines = load 'hdfs://myPoints.txt';
points = map lines -> kmeans.parsePoints();
centroids = load 'hdfs://myInitialCentroids.txt';
final_centroids = repeat centroids AS current_centroid for 50 { 
distance = map points -> kmeans.selectNearestCentroid() with broadcast current_centroid; 
centroids_sum = reduce distance -> kmeans.reduce();
new_centroids = map centroids_sum -> kmeans.average(); }
store final_centroids 'hdfs:///output/kmeans';
\end{lstlisting}

\begin{lstlisting}[basicstyle=\scriptsize\normalfont\sffamily,aboveskip=0pt,belowskip=0pt,float=h!,label=list:kmeanplan,numbers=none,numbers=left,caption=K-means task using the Scala API.,language=scala]
val points = plan.readTextFile("hdfs://myPoints.csv")
	.map(createPoints)
val initialCentroids = plan.loadCollection(Kmeans.createRandomCentroids(k))
val finalCentroids = initialCentroids.repeat(iterations, { currentCentroids =>
	val newCentroids = points.mapJava(
		new SelectNearestCentroid,
	)
	.withBroadcast(currentCentroids, "centroids")
	.reduceByKey(_.centroidId,  + )
	.map(_.average newCentroids})
finalCentroids.collect()
\end{lstlisting}

{\em PolyJoin} is a common task in polystore scenarios, \ie~joining several datasets from different data sources.
In this case, we consider the TPC-H Q5 and assume that:
the \at{region}, \at{suppliers}, and \at{customer} relations are on Postgres;
the \at{nations} relations is on the local file system; and
the \at{orders} and \at{lineitem} relations are on HDFS.
Despite the complexity of this query, we observe that the RheemLatin query (Listing~\ref{list:polylatin}) and the Scala (Listing~\ref{list:polyplan}) are still simple as they follow the logical flow of the task itself. Lines~1-7 in Listing~\ref{list:polylatin} load the dataset, Lines~8-12 select and project the required tuples, and Lines~13-22 join the resulted tuples before making the group-by in Line~23.

\begin{lstlisting}[basicstyle=\scriptsize\normalfont\sffamily,aboveskip=0pt,belowskip=0pt,float=h!,label=list:polylatin,numbers=none,numbers=left,caption=PolyJoin task in RheemLatin.,language=rheemlatin]
import '/polyjoin/udfs.class' as polyjoin;
region = load 'postgres:///tpch/region';
suppliers = load 'postgres:///tpch/suppliers';
customers = load 'postgres:///tpch/customers';
nations = load 'file:///nations' delimiter '|';
orders = load 'hdfs:///orders' delimiter '|';
lineitems = load 'hdfs:///lineitems' delimiter '|';
region_filter = filter region[1] ==  'ASIA';
region_project = map region_filter -> { polyjoin.projectRecord(0, 1) };
suppliers_project = map suppliers -> { polyjoin.projectRecord(0, 3) };
customers_project = map customers -> { polyjoin.projectRecord(0, 3) };
order_filter = filter orders -> { polyjoin.isBetween( 4, '1994-01-01', '1995-01-01') };
join1 = join nation[2], region_project[0];
map_join1 = map join1 -> { polyjoin.tuple2Record(0, 0, 0, 1) };
join2 = join map_join1[0], customers_project[1];
map_join2 = map join2 -> { polyjoin.tuple2Record(0, 0, 0, 1, 1, 0) };
join3 = join map_join2[2], order_filter[0];
map_join3 = map join3 -> { polyjoin.tuple2Record(0, 0, 0, 1, 1, 0) };
join4 = join map_join3[2], lineitems[0];
map_join4 = map join4 -> { polyjoin.tuple2Record(0, 0, 0, 1, 1, 2, 1, 5, 1, 6) };
join5 = join map_join4 -> { polyjoin.record2Tuple(2, 0) }, suppliers_project -> { polyjoin.record2Tuple(0, 1) };
map_join5 = map join5 -> { polyjoin.tuple2Record(0, 1, 0, 3, 0, 4) }; 
groupBy = groupby map_join5[0];
store groupBy '/output/polyjoin';
\end{lstlisting}

\begin{lstlisting}[basicstyle=\scriptsize\normalfont\sffamily,aboveskip=0pt,belowskip=0pt,float=h!,label=list:polyplan,numbers=none,numbers=left,caption=PolyJoin task using the Scala API.,language=scala]
val regions: DataQuanta[Record] = plan.readTable("postgres:///tpch/region")
	.map( createRecord(_))
	.filter((r: Record) => r.getString(1) == "ASIA")
	.map( projectRecord(_, 0, 1))
val suppliers: DataQuanta[Record] = plan.readTable("postgres:///tpch/supplier")
	.map[Record]( createRecord(_))
	.map[Record]( projectRecord(_, 0, 3))
val customers: DataQuanta[Record] = plan.readTable("postgres:///tpch/customer")
	.map[Record]( createRecord(_))
	.map( projectRecord(_, 0, 3))
val nations: DataQuanta[Record] = plan.readTextFile("file:///nation")
	.map( createRecord(_))
val orders: DataQuanta[Record] = plan.readTextFile("hdfs:///order")
	.map( createRecord(_))
	.filter( isBetween(_, 4, fromData, toDate) )
val lineitems: DataQuanta[Record] = plan.readTextFile("hdfs:///lineitem")
	.map(createRecord(_))
nations
	.join( getColumn(, 2), regions, getColumn(, 0))
	.map( tuple2Record(_, 0, 0, 0, 1))
	.join( getColumn(, 0), customers, getColumn(, 1))
	.map( tuple2Record(_, 0, 0, 0, 1, 1, 0))
	.join( getColumn(, 2), orders, getColumn(, 1))
	.map( tuple2Record(_, 0, 0, 0, 1, 1, 0))
	.join( getColumn(, 2), lineitems, getColumn(, 0))
	.map( tuple2Record(_, 0, 0, 0, 1, 1, 2, 1, 5, 1, 6))
	.join[Record, Tuple2[String, String]]( record2Tuple(, 2, 0), suppliers, record2Tuple(, 0, 1))
	.map( tuple2Record(_, 0, 1, 0, 3, 0, 4))
	.groupByKey((r: Record) => r.getField(0))
	.collect()
\end{lstlisting}

%% file: musketeer.tex
We experimentally compare \rheem with its closest competitor, Musketeer~\cite{gog2015musketeer}. More experiments concerning the optimizer can be found in~\cite{rheem-optimizer-arxiv}.

\myparagraph{Setup}
We ran our experiments on a cluster of 10 machines.
Each node has one $2$\,GHz Quad Core Xeon processor, $32$\,GB main memory, $500$\,GB SATA hard disks, a $1$\,Gigabit network card and runs $64$-bit platform Linux Ubuntu 14\@.04\@.05.
In \rheem we used the following platforms:
Java's Stream library (\textsf{JavaStreams}),
Spark~1.6.0 (\textsf{Spark}),
Flink~1.3.2 (\textsf{Flink}),
GraphX~1.6.0 (\textsf{GraphX}),
Giraph~1.2.0 (\textsf{Giraph}),
a Java graph library (\textsf{JGraph}), and
HDFS~2.6.0 to store files.
We used all these platforms with their default settings and configured the maximum RAM of each platform to $20$\,GB.
We disabled the \rheem stage parallelization feature to have only one single platform running at any time.
We obtained all the cost functions required by our optimizer as described in Section~\ref{section:core_costlearner}.
We considered the cross-community pagerank task (\task{CrocoPR}), because the authors reported this task to be a case where Musketeer chooses multiple platforms.
Note that, for fairness reasons, we perform the data preparation part of \task{CrocoPR} (\ie~union the different communities pages) as a separate script for Musketeer.
This is because its language (Mindi) is not optimized for dealing with UDFs, thereby it would be much slower to provide the data preparation as a UDF.
In contrast, \rheem seamlessly performs both parts (data preparation and page rank) as a single task.
We used the DBPedia pagelinks dataset ($20$ GB).

\begin{figure}[!t]
	\centering
	\includegraphics[width=\columnwidth]{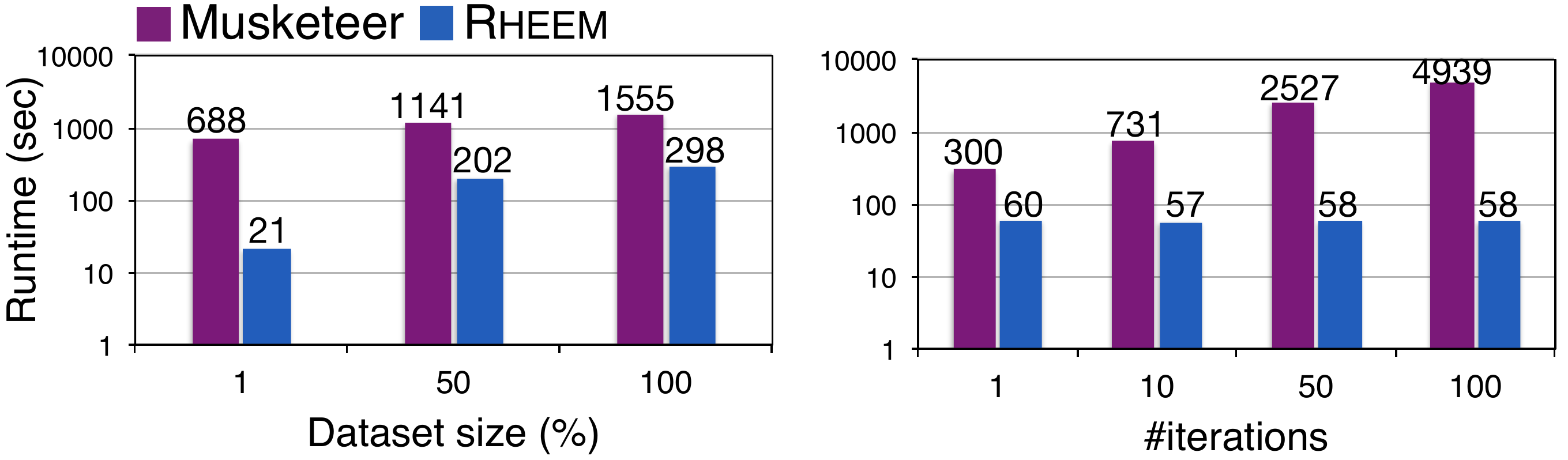}
	\vspace{-0.5cm}
	\caption{\rheem outperforms Musketeer by more than one order of magnitude.}
	\label{figure:comparison}
	\vspace{-0.5cm}
\end{figure}

\myparagraph{Results}
Figure~\ref{figure:comparison} shows the results in log scale when varying the dataset sizes for $10$ iterations and the number of iterations for $10\%$ of the dataset.
Overall, we observe the superiority of \rheem over Musketeer, especially as the number of iterations increases: \rheem is up to $85$ times faster than Musketeer.
Note that, in contrast to Musketeer, \rheem keeps its runtime constant as the number of iterations increases.
This is because:
(i)~Musketeer, among other things, checks dependencies, compiles and package the code, and writes the output to HDFS at each iteration (or stage), which comes with a high overhead;
~(ii)~\rheem executes the page rank part of the task (\ie~after the data preparation) on \pl{JavaStreams}, which allows it to perform each iteration with almost zero overhead.

%% file: limitations.tex

As of now, \rheem does not support any stream processing platforms.
While users can easily supply new batch processing platforms, stream processing requires to extend \rheem's core.
We plan to do so by following the lambda architecture paradigm~\cite{marz2015big}.
In addition, \rheem currently relies on the fault-tolerance of the underlying platforms and is, thus, susceptible to failures while moving data across platforms. We plan to incorporate some basic fault-tolerance mechanism at the cross-platform level.
Other remaining issues include:
adding methods that speed up inter-platform communications, such as the one proposed in~\cite{pipegen}, integrating \rheem with resource managers to incorporate changes in the availability of computing resources, and supporting simultaneous execution of \rheem jobs.

%% file: relatedwork.tex

The research and industry communities have proposed a myriad of different data processing platforms~\cite{MapReduce-ACM,hbase,apacheSpark,PostgreSQL,husky,stratosphere}.
In contrast, we do not provide a data processing platform but a novel system on top of them.

Cross-platform data processing has been in the spotlight only very recently.
Some works focus only on integrating different data processing platforms with the goal of alleviating users from their intricacies~\cite{apacheDrill,apacheBeam,luigi,prestoDB,bigdawg-demo}.
However, they still require expertise from users to decide when to use a specific data processing platform.
For example, BigDAWG~\cite{bigdawg-demo} requires users to specify where to run tasks via its \op{Scope} and \op{Cast} commands, which already require expertise from users.
Only few works share a similar goal with us~\cite{gog2015musketeer,SimitsisWCD12,ires-bigdata,dbms+,myria}.
However, they substantially differ from \rheem.
Two main differences are that they consider neither data movement costs nor progressive task optimization techniques, although both aspects are crucial in cross-platform settings.
Additionally, each of these works differs from \rheem in various ways.
As Musketeer's main goal is to decouple front-end languages (\eg~SQL and PigLatin) from the underlying platforms~\cite{gog2015musketeer}, it is not as expressive and extensible as \rheem.
Furthermore, as it maps task patterns to specific underlying platforms, it is not clear how one can efficiently map a task when having similar platforms (\eg~Spark vs. Flink or Postgres vs. MySQL).
Similarly, in Myria~\cite{myria}, it is hard to allocate tasks when having similar platforms because it comes with a rule-based optimizer.
Additionally, its rule-based optimizer also makes it hard to maintain.
IReS~\cite{ires-bigdata} supports only 1-to-1 mappings between abstract tasks and their implementations, which limits expressiveness and optimization opportunities.
Moreover, it assumes direct data movement paths between platforms, which is hard to maintain for many platforms.
QoX focuses only on ETL workloads~\cite{SimitsisWCD12}.
DBMS+~\cite{dbms+} is limited by the expressiveness of its declarative language and hence it is neither adaptive nor extensible.
Other complementary works focus on improving data movement across different platforms~\cite{pipegen} or libraries by using a common intermediate representation and executing the scripts in LLVM~\cite{weld}, but none of them address the cross-platform optimization problem.
Tensorflow~\cite{tensorflow2015-whitepaper} follows a similar idea, but for cross-device execution of machine learning tasks and thus it is orthogonal to \rheem.
In fact, \rheem could use TensorFlow as an underlying platform.

The research community has also studied the problem of federating relational databases~\cite{federatedDBsSurvey}.
Garlic~\cite{garlic}, TSIMMIS~\cite{tsimmis}, and InterBase~\cite{interbase} are just three examples.
However, all these works significantly differ from \rheem in that they consider a single data model and simply push query processing to where the data is.
Other works integrate Hadoop with an RDBMS~\cite{miso,polybase}, however, 
one cannot easily extend them to deal with more diverse tasks and platforms.

%% file: conclusion.tex

Given today's data analytic ecosystem, supporting cross-platform data processing has become rather crucial in organizations.
We have identified four different situations in which an application requires or benefits from cross-platform data processing.
Driven by these cases, we built \rheem, a cross-platform system that decouples applications from data processing platforms to achieve efficient task execution over multiple platforms. 
\rheem follows a cost-based optimization approach for splitting an input task into subtasks and assigning each subtask to a specific platform, such that the cost (\eg~runtime or monetary cost) is minimized.
Our experience while building \rheem raised several interesting questions that need to be addressed in the future, namely:
{\em How can we (i)~reduce the inter-platform data movement costs?
(ii)~address the cardinality and cost estimation problem?
(iii)~efficiently support fault-tolerance across platforms?
(iv)~add new platforms automatically? and
(v)~improve data exploration in cross-platform settings?}